\documentclass[aps,bibnotes,twocolumn,showpacs, showkeys, superscriptaddress,preprintnumbers,amsmath,amssymb,lengthcheck,pra]{revtex4-1}

\usepackage{graphicx}
\usepackage{dcolumn}
\usepackage{bm}
\usepackage{color}
\usepackage{soul}

\newcommand{\bc}{\textcolor{black}}

\begin{document}

\title{Optical Nonlinear Dark X-Waves}

\author{Fabio Baronio} 
\email{fabio.baronio@unibs.it}
\author{Stefan Wabnitz} 
\affiliation{INO CNR and Dipartimento di Ingegneria dell'Informazione, Universit\`a di Brescia, Via Branze 38, 25123 Brescia, Italy}

\author{Shihua Chen}
\affiliation{Department of Physics, Southeast University, Nanjing 211189,China}

\author{Miguel Onorato}
\affiliation{Dipartimento di  Fisica, Universit\`a di Torino, Via P. Giuria 1, 10125 Torino, Italy}
\affiliation{Istituto Nazionale di Fisica Nucleare, INFN, Sezione di Torino, 10125 Torino, Italy}

\author{Stefano Trillo}
\affiliation{Dipartimento di  Ingegneria, Universit\`a di Ferrara, Via Saragat 1, 44122 Ferrara, Italy}

\author{Yuji Kodama} 
\affiliation{Department of Mathematics, Ohio State University, Columbus, OH 43210, USA}

\begin{abstract}
We introduce spatiotemporal optical dark X solitary waves of the (2+1)D {hyperbolic} nonlinear Schr\"odinger equation (NLSE), that rules wave propagation in a self-focusing and normally dispersive medium. 
These analytical solutions are derived by exploiting the connection between such NLSE and a well known equation of hydrodynamics, namely the type II Kadomtsev-Petviashvili (KP-II) equation.
As a result, families of shallow water X soliton solutions of the KP-II equation are mapped into optical dark X solitary wave solutions of the NLSE. Numerical simulations show that optical dark X solitary waves may propagate for long distances (tens of nonlinear lengths) before they eventually break up, owing to the modulation instability of the continuous wave background. This finding opens a novel path for the excitation and control of X solitary waves in nonlinear optics. 
\end{abstract}


\maketitle

Laser pulse and beam shaping techniques \cite{wein09}, and in particular the possibility to obtain localized distorsionless (both non-diffractive and non-dispersive) wave packets \cite{reca08} are of paramount importance in many fields of basic and applied research such as atomic physics, spectroscopy, communications, and medicine. 
In this context, X waves, originally discovered in acoustic \cite{lu92}, have been established as a new paradigm in areas ranging from classical to quantum optics \cite{saari97,conti03,ditra03,kolesic04,faccio06,ciat07,silb07,orni15}. 
Specifically, envelope X waves emerged to be the key for understanding the dynamics in bi-dispersive settings characterized by opposite signs of dispersions, the most natural situation being the spatio-temporal dynamics ruled by standard paraxial diffraction in normally dispersive media. They exist in the linear regime \cite{reca08,porra03,efre04}, being non-monochromatic superpositions of non-diffracting modes (Bessel \cite{Bessel} or cosine modes in transverse 2D and 1D, respectively). However, it is their {\it nonlinear} counterpart (obtained via numerical dressing of linear solutions \cite{conti03,SFbook}) that attracted much interest because of the capability to emerge spontaneously in different experiments concerning parametric converters \cite{conti03,ditra03}, Kerr media \cite{kolesic04, faccio06}, and periodic structures \cite{silb07}.\\
\indent The nonlinear regime, however, poses a number of challenges that remained unaddressed to date. First, there are no available methods to construct analytical solutions.
Exact nonlinear X wave solutions are known only in the presence of a potential which rule out the most interesting experimental situations that involve free propagation \cite{efre09}.
Second, only bright nonlinear X waves have been reported, whereas the possibility to find dark X waves over a finite background was overlooked (in the linear regime, the validity of the superposition principle guarantees that such dark X waves trivially exist as the sum of the background, which is solution of the homogeneous wave equation, and the sign reversed version of any bright X wave solution, which is still a solution, due to invariance of the wave equation under sign inversion).\\  
\indent In this Letter, we show that both of such restrictions can be overcome at once by exploiting a transformation \cite{turi88,peli95,baro16} that maps the most universal of bidispersive nonlinear models, namely the (2+1)D nonlinear Schr\"odinger equation (NLSE) into the (2+1)D Kadomtsev-Petviashvili (KP) equation. 
\bc{The latter constitutes the natural extension of the well known (1+1)D Korteweg-de Vries (KdV) equation and is widely employed in plasma and hydrodynamics (see e.g.\cite{KP70,mil77,as81,kod04,kod10,kod11}) in its two different forms, the so-called KP-I type and KP-II type, depending on the sign of the transverse perturbation to the KdV equation.}
In particular, we first show how the original two-soliton X-shaped solutions \cite{kod04} of the KP-II generate nonlinear dark X solitary solutions of the hyperbolic NLSE,
which are potentially observable in the regime investigated experimentally in \cite{ditra03,faccio06,silb07}.
Then, we exploit a different family of X soliton solution of the KP-II equation, namely the \textit{Toda-type} \cite{kod04}, and find their optical dark X solitary counterpart of the NLSE.\\
\indent In the presence of group-velocity dispersion and one dimensional diffraction, the dimensionless time-dependent paraxial wave equation in cubic Kerr media reads as \cite{conti03}:
\begin{equation}\label{2DNLS}
iu_z+\frac{\alpha}{2}u_{tt}+\frac{\beta}{2}u_{yy}+\gamma |u|^2 u=0,
\end{equation}
namely the  (2+1)D, or more precisely (1+1+1)D, NLSE, where $u(t,y,z)$ stands for the complex wave envelope, and $t,y$ represent the retarded time (in the frame traveling at the natural group-velocity) and the spatial transverse coordinate, respectively, and $z$ is the longitudinal propagation coordinate. Each subscripted variable in Eq. (\ref{2DNLS}) stands for partial differentiation. $\alpha, \beta, \gamma$ are normalized real 
constants that describe the effect of dispersion, diffraction and Kerr nonlinearity, respectively. 

We refer \eqref{2DNLS} to as \emph{elliptic} NLSE if $\alpha\beta>0$, and \emph{hyperbolic} NLSE if $\alpha\beta<0$.
In the case of weak nonlinearity, weak diffraction and slow modulation, the dynamics of optical NLSE dark envelopes
 $u(t,y,z)$ may be related to the hydrodynamic KP variable $\eta(\tau,\upsilon,\varsigma)$ as follows \cite{baro16}:
\begin{align}\label{NLSEKP}
u(t,y,z)\simeq \sqrt{\rho_0+\eta(\tau,\upsilon,\varsigma)} \ \ e^{i[\gamma\rho_0z -(\gamma/c_0) \int  \eta(\tau,\upsilon,\varsigma) d\tau ]},
\end{align}
where $\rho_0$ stands for a background continuous wave amplitude,   $\eta(\tau,\upsilon,\varsigma)$ represents a small amplitude variation, say $\eta\sim \mathcal{O}(\epsilon)$ with $0<\epsilon\ll 1$ and the order one background $\rho_0$; 
$\eta(\tau,\upsilon,\varsigma)$ satisfies the KP equation,
\begin{equation}\label{KP}
\left(-\eta_\varsigma+\frac{3\alpha\gamma}{2c_0}\eta\eta_\tau+\frac{\alpha^2}{8c_0}\eta_{\tau\tau\tau}\right)_\tau-\frac{c_0 \beta}{2\alpha}\eta_{\upsilon\upsilon}=0,
\end{equation}
where $\tau=t-c_0z, \upsilon=y$, $\varsigma=z$ with  $c_0=\sqrt{-\gamma\alpha\rho_0}$, $\alpha \gamma<0$
(see  \cite{baro16} for further details).
 
\bc{Contrary to the case dealt with in \cite{baro16}, which considered lump solutions of the elliptic NLSE for defocusing media ($\alpha, \beta>0, \gamma<0$), derived through Eq. (\ref{NLSEKP}) from the KP-I equation [Eq. (\ref{KP}) with $\beta/\alpha>0$], in this letter we focus our attention on the combined action of diffraction and normal dispersion for self-focusing media. The latter case corresponds to $\alpha <0$ and $\beta, \gamma>0$, i.e. to the focusing hyperbolic NLSE 
linked through Eq. (\ref{NLSEKP}) to the KP-II equation [$\beta/\alpha<0$ in Eq. (\ref{KP})].
Interestingly enough, the results that we derive below have relevance also for different contexts where the same hyperbolic NLSE applies, such as the propagation in suitably engineered lattices giving rise to effective negative diffraction \cite{silb07,eise00}, or the dynamics of envelope water waves \cite{yuen80} (in both cases $t$ represents an additional spatial variable).
}

In order to proceed further we fix, without loss of generality, the coefficients of Eq. (\ref{2DNLS}) as $\alpha=-4 \sqrt{2}, \beta= 6\sqrt{2}, \gamma=2\sqrt{2}$, and the background to unit $\rho_0=1$. On one hand, this allows to cast Eq. (\ref{KP}) in its standard KP-II form $(-\eta_{{\varsigma}}-6 \eta\eta_\tau+\eta_{\tau\tau\tau})_\tau+3\eta_{\upsilon\upsilon}=0$.
\bc{On the other hand, this fixes the scaling between the dimensionless variables $z,t,y$ in Eq. (\ref{2DNLS}) and the corresponding real-world quantities $Z=Z_0z,T=T_0t,Y=Y_0y$. 
The longitudinal scaling factor turns out to be $Z_0=2\sqrt{2} L_{nl}$, where $L_{nl}=(\gamma_{phys} I_0)^{-1}$  is the usual nonlinear length associated with the intensity $I_0$ of the background and $\gamma_{phys}=k_0 n_{2I}$, $n_{2I}$ being the Kerr nonlinear index and $k_0$ the vacuum wavenumber. The "transverse" scales read as $T_0=\sqrt{k'' L_{nl}/2}$ and $Y_0=\sqrt{L_{nl}/(3k_0 n)}$, where $k''$ and $n$ are the group-velocity dispersion and the linear refractive index, respectively.}

\begin{figure}[h!]
\begin{center}
\includegraphics[width=7.7cm]{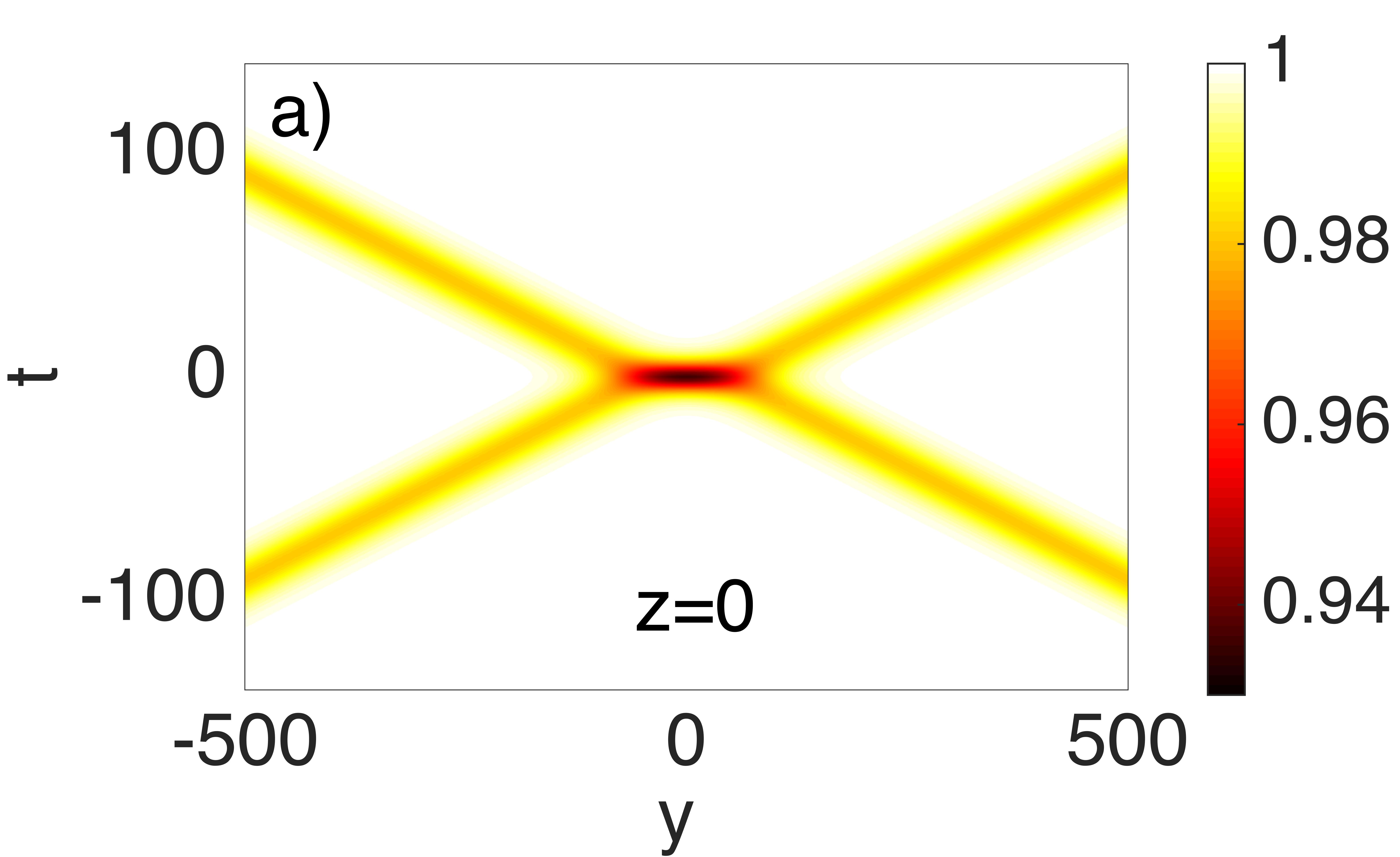}
\includegraphics[width=7.7cm]{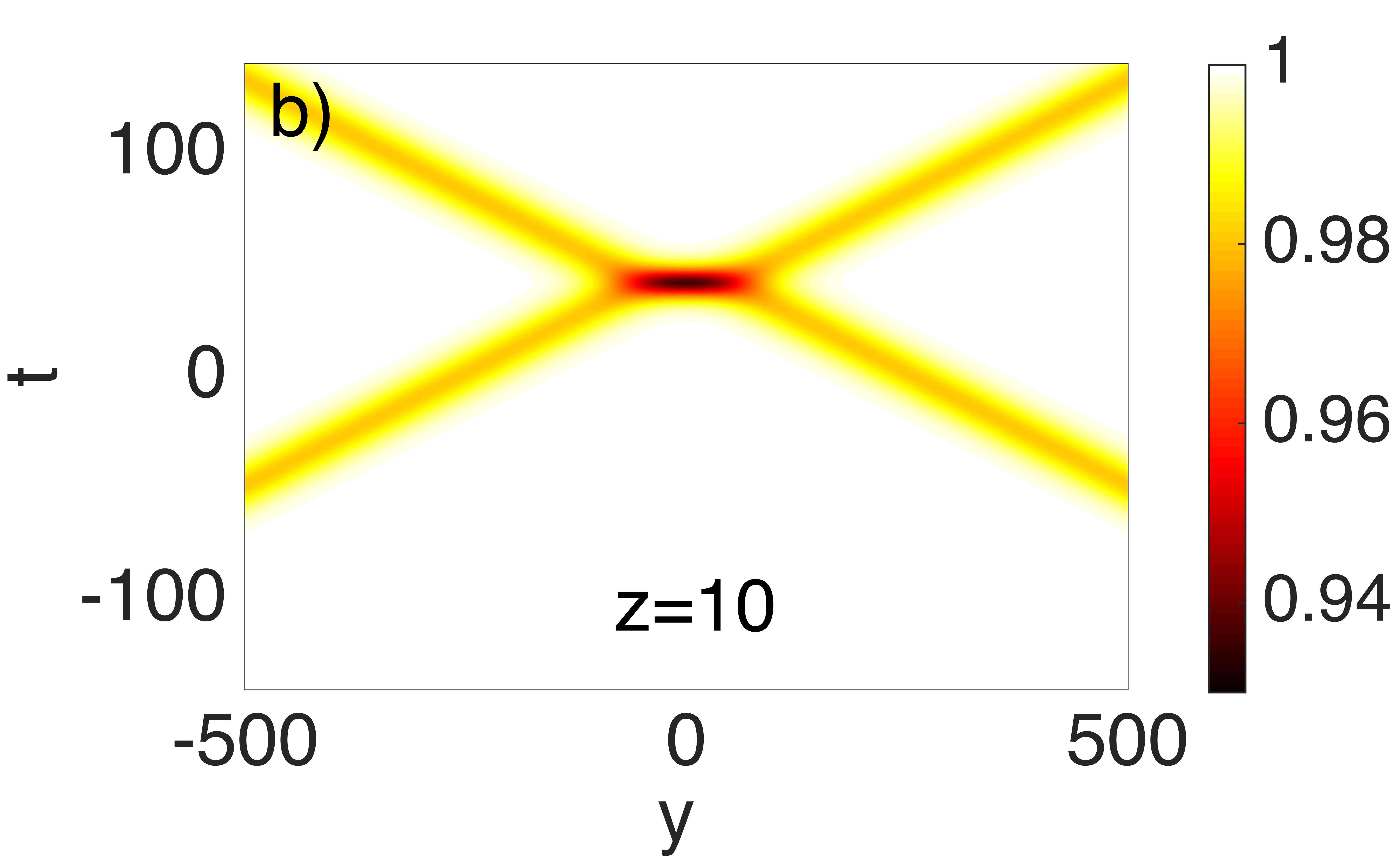}
\includegraphics[width=7.7cm]{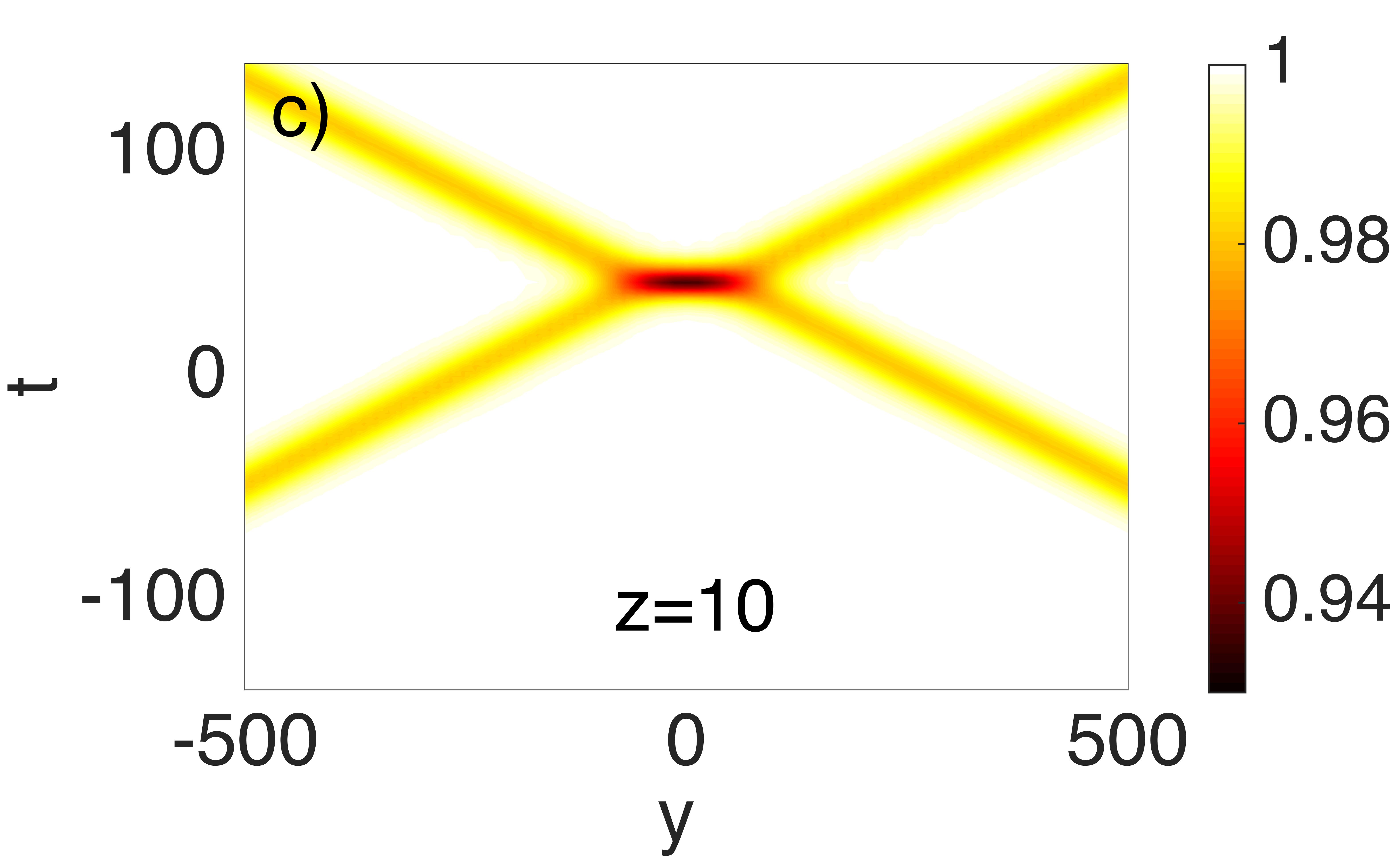}
\caption{\label{f2} Spatio-temporal NLSE envelope intensity distribution $|u|^2$, in the $(y,t)$ plane, showing a typical dark X solitary wave.
a) $|u|^2$ at $z=0$ and b) at $z=10$. c) numerically computed profile $|u|^2$ according to hyperbolic NLSE (1), at the propagation distance $z=10$. 
Here, $\epsilon_1=0.2$, $\epsilon_2=0.001$. 
}
\end{center}
\end{figure}

The KP-II equation admits complicated soliton solutions, mostly discovered in the last decade, 
which may describe non-trivial web patterns generated under resonances of line-solitons observable in shallow water \cite{kod04,kod10,kod11}.
Among these, we first consider the so-called \textit{O-type} bright X-shaped 2-soliton solution of the KP-II (note that the name \emph{O-type} is due to 
the fact that this solution was \emph{originally} found by the Hirota bilinear method \cite{kod04}; it should not be confused with conical $O$-waves of the elliptic NLSE \cite{reca08},  which have no relevance in the present context).
When considering small amplitude regimes, a formula for an exact  \textit{O-type} solution of Eq. (\ref{KP}) can be expressed as follows \cite{kod04},
\begin{equation}\label{2kps}
\eta (\tau, \upsilon, \varsigma) = - 2 \left( \ln F\right)_{\tau\tau}, 
\end{equation}
where the function $F(\tau,\upsilon,\varsigma)$ is given by $F=f_1+f_2$ with
\begin{gather*}\label{def}
f_1= (\epsilon_1 + \epsilon_2) \, {\rm cosh}[  (\epsilon_1 - \epsilon_2) \tau+4 \,(\epsilon_1^3 - \epsilon_2^3) \varsigma] \\
 f_2=2 \sqrt{\epsilon_1 \epsilon_2} \,{\rm cosh}[ (\epsilon_1^2 - \epsilon_2^2)\upsilon].
\end{gather*}
$\epsilon_1, \epsilon_2$ are small real positive parameters which are related to the amplitude, width and the angle of the \textit{O-type} X-soliton solutions.

The existence of (2+1)D NLSE dark X solitary waves $u(t,y,z)$, is directly given through Eq. (\ref{NLSEKP}),
exploiting the soliton expression for $\eta(\tau,\upsilon,\varsigma)$  in Eq. (\ref{2kps}).
%

%
Figure \ref{f2} shows the spatiotemporal envelope intensity profile
$|u|^2$ of a (2+1)D NLSE dark X solitary wave of the hyperbolic NLSE. The solution is shown in the $(y,t)$ plane, 
at $z=0$ in Fig. \ref{f2}a and at $z=10$ in Fig. \ref{f2}b.
In this particular example we have chosen $\epsilon_1=0.2$, $\epsilon_2=0.001$. 
Specifically, Figs. \ref{f2}a,b illustrate a solitary solution which describes the X-interaction of multiple dark line solitons. 
Asymptotically, the solution reduces to two line dark waves for $t\ll0$ and two for $t\gg0$, 
with intensity dips $\frac{1}{2}(\epsilon_1-\epsilon_2)^2$ and characteristic angles $\pm{\rm tan}^{-1}(\epsilon_1+\epsilon_2)$, 
measured  from the $y$ axis. The maximum value of the dip in the interaction region is
$2 (\epsilon_1-\epsilon_2)^2  \, (\epsilon_1+\epsilon_2) / (\epsilon_1+\epsilon_2 +2 \sqrt{\epsilon_1 \epsilon_2})$.

Next, we numerically verified the accuracy and stability of the analytically predicted \textit{O-type} dark X solitary wave of the NLSE. 
To this end, we made use of a standard split-step Fourier technique, commonly adopted in the numerical solution of the NLSE Eq. (\ref{2DNLS}).
We took the dark wave envelope at $z=0$ as the numerical input: $u(t,y,z=0) = \sqrt{1+\eta(\tau=t,\upsilon=y,\varsigma=0)} \exp{\left[i \phi(\tau=t,\upsilon=y,\varsigma=0) \right]}$,
where $\eta$ is the X-soliton solution (\ref{2kps}) and $ \phi=-(\gamma/c_0) \int_\tau \eta$.
Fig. \ref{f2}c shows the $(y,t)$ profile of the numerical solution of the hyperbolic NLSE obtained at $z=10$,
which shows excellent agreement with the analytical solution from Eq. (\ref{2kps}) computed at $z=10$, and reported in Fig. \ref{f2}b.
\bc{These results prove that the proposed solutions propagate as X-shaped nonlinear invariant modes of the NLSE, being subject only to a net delay due to the deviation $c_0$ from the natural group-velocity of the medium.
The spatio-temporal Fourier spectrum of these waves is also X-shaped (result not shown). These features allow us to classify such modes in the broad class of diffraction-free and dispersion-free X waves.
It is worth pointing out, however, that there are important differences with the more general nonlinear X wave solution reported in the literature for the (3+1)D hyperbolic NLSE, i.e. for 2D diffraction \cite{conti03,SFbook}. 
In particular, the latter type of X waves exhibit a characteristic decay $1/r$ along the spatial coordinate $r$ which is characteristic of Bessel functions constituting the building blocks of X waves in the linear limit.
Conversely, in the present case, the dark X solitary waves have, by construction, constant asymptotic (i.e. the line solitons), while the transformation in Eq. (\ref{NLSEKP}) becomes meaningless in the linear limit. 
Nevertheless, the asymptotic state is compatible with 1D transverse diffraction (see, e.g. Fig. 1 in \cite{silb07}), a regime where the connections between the linear and nonlinear X-waves have not been exhaustively investigated yet.
Of course any finite energy realisation of the present type of solutions should consider a spatio-temporal  envelope modulation of the X solitary wave that decays to zero sufficiently slow in $(t,y)$ compared with the extension of the solitary central notch, similarly to the case of dark solitons in (1+1)D \cite{dark88}.}

\begin{figure}[h!]
\begin{center}
\includegraphics[width=7.7cm]{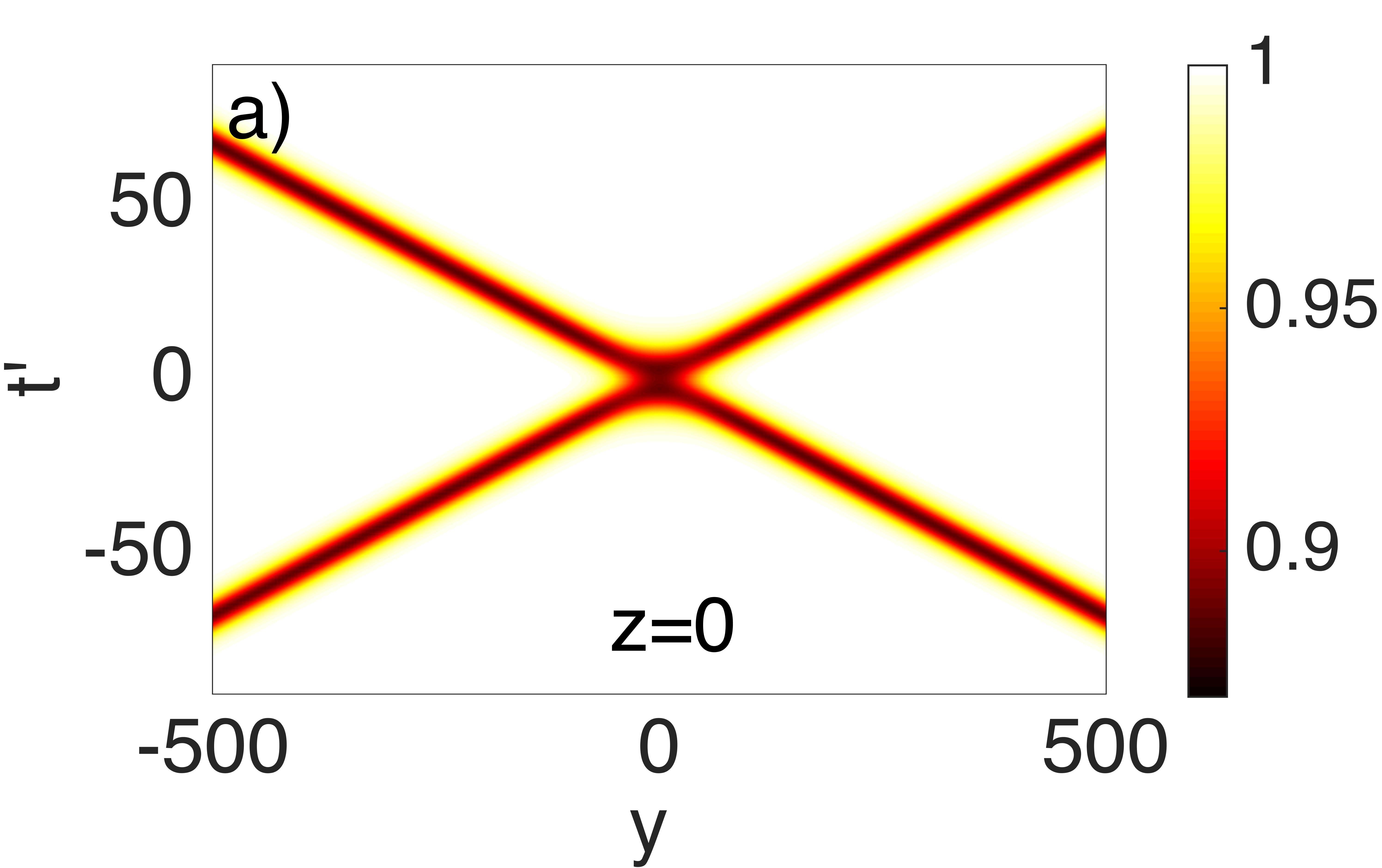}
\includegraphics[width=7.7cm]{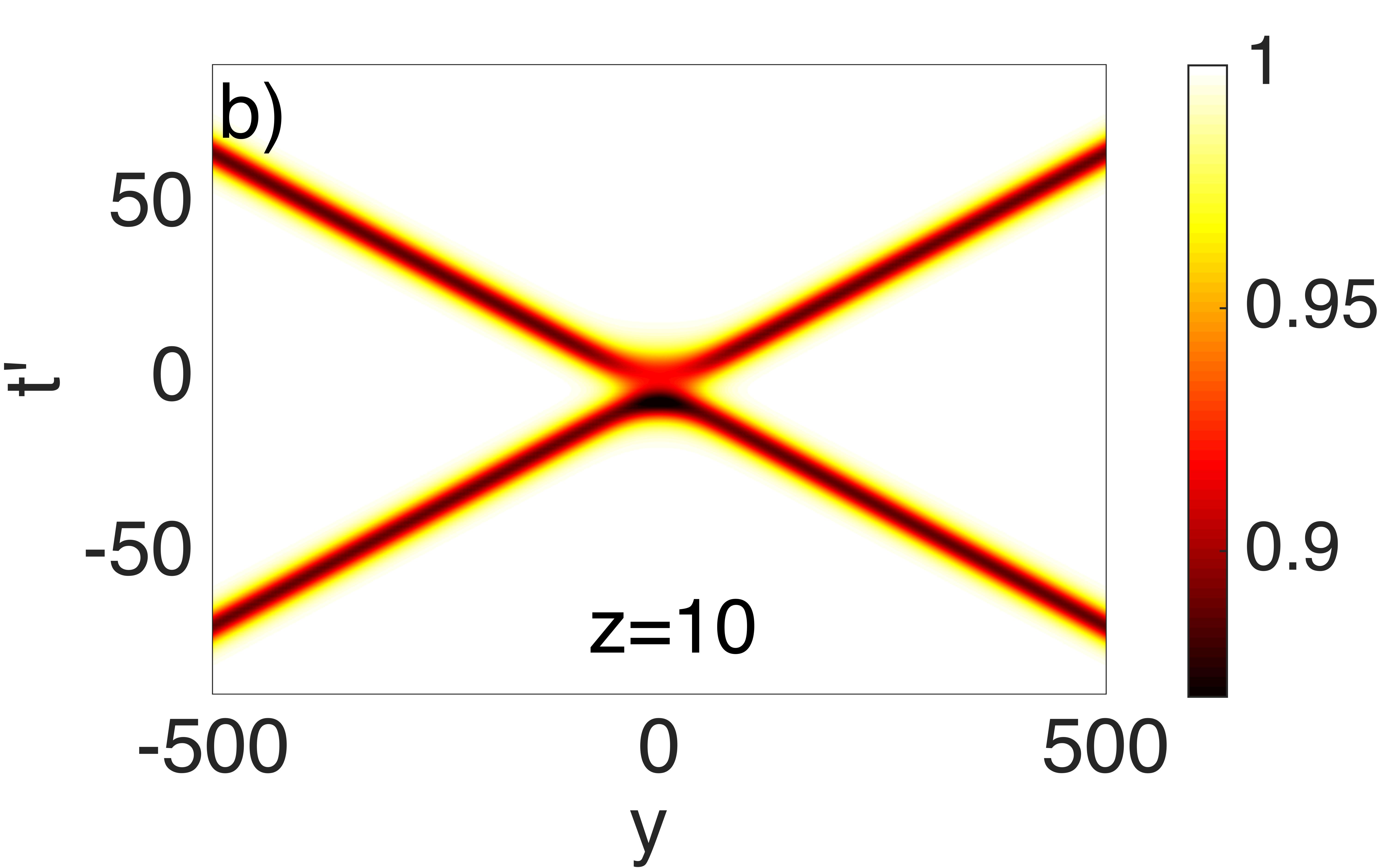}
\includegraphics[width=7.7cm]{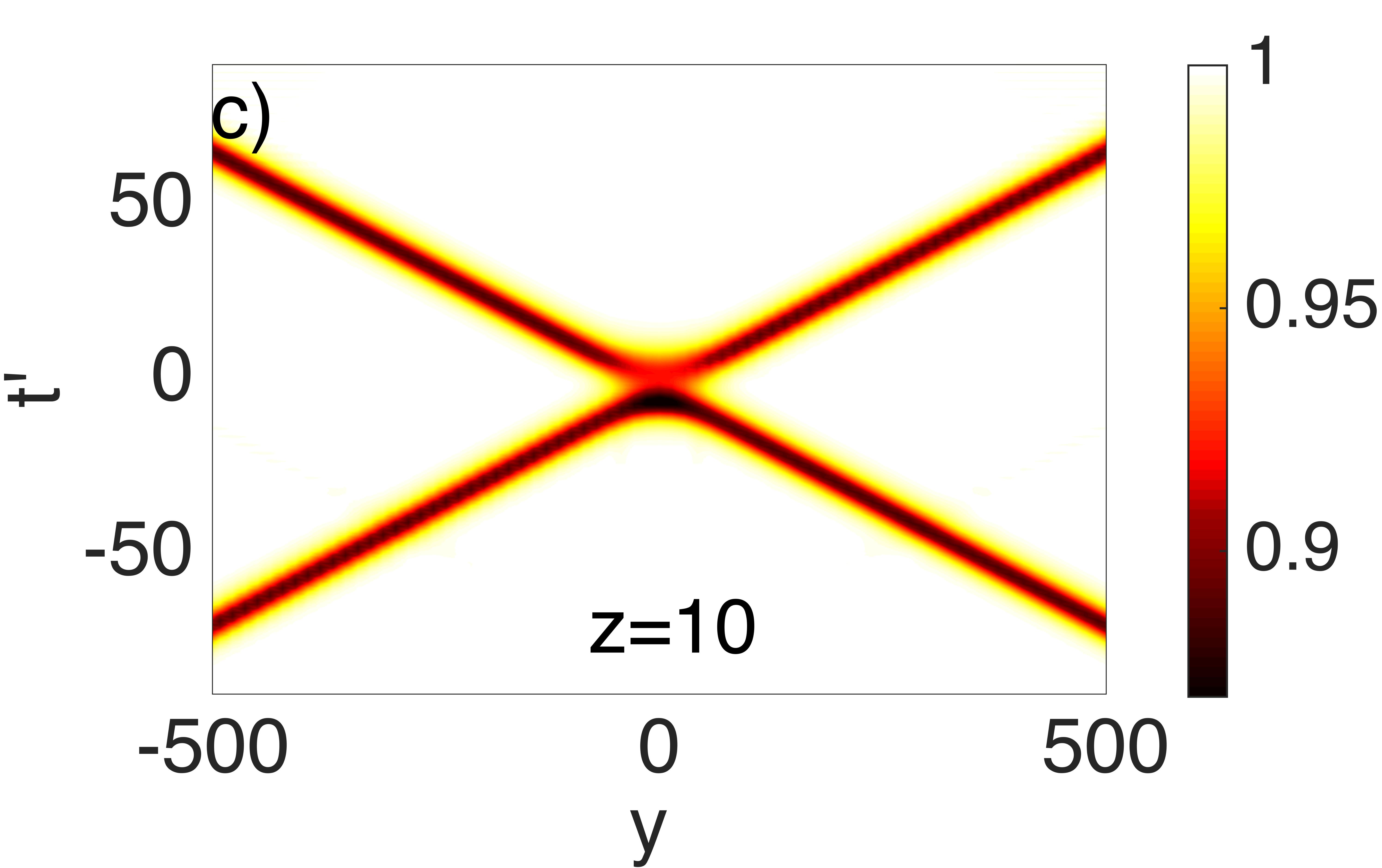}
\caption{\label{f3} Spatio-temporal NLSE envelope intensity distribution
$|u|^2$, in the $(y,t')$ plane ($t'=t-c_0z$), showing a  dark X solitary wave fission.
a) $|u|^2$ at $z=0$ and b) at $z=10$. c) numerically computed profile $|u|^2$ at 
the propagation distance $z=10$.
Here, $\epsilon_1=0.31$, $\epsilon_2=0.17$. 
}
\end{center}
\end{figure}

\bc{
The link between the hyperbolic NLSE and the KP-II equation is not limited to the type of invariant waves discussed above.
Among the variety of other types of KP-II X-shaped soliton solutions found in the last decade, e.g. so-called \textit{T-type} and \textit{P-type} soliton solutions \cite{kod04},
below we discuss the relevance of the \textit{T-type} solitons. They originate from soliton solutions found in the \emph{Toda lattice} equation \cite{bik03}, and describe a fully resonant interaction of two line-solitons.
}
When considering the small amplitude regime, exact  \textit{Toda-type} soliton solution of Eq. (\ref{KP}) can be expressed in the form
\begin{equation}\label{2kpsT}
\eta (\tau, \upsilon, \varsigma) = - 2 \left(\ln F\right)_{\tau\tau}, 
\end{equation}

where $F=f_1+f_2$ with
\begin{gather*}\label{def}
f_1= (\epsilon_1 + \epsilon_2)^2 \, {\rm cosh}[  (\epsilon_1 - \epsilon_2) \tau+4 \,(\epsilon_1^3 - \epsilon_2^3) \varsigma] \\
 f_2=2 (\epsilon_1- \epsilon_2) \sqrt{\epsilon_1 \epsilon_2} \,  \{ {\rm cosh}[ (\epsilon_1^2 - \epsilon_2^2)\upsilon]  + \\
   {\rm cosh}[ 4 \, (\epsilon_1^3 + \epsilon_2^3)\varsigma +  (\epsilon_1 + \epsilon_2) \tau ] \}.
\end{gather*}
$\epsilon_1, \epsilon_2$ are small real positive parameters
which rule the amplitude, width and the angle of the \textit{Toda-type} solution.

Again, the existence of (2+1)D NLSE dark X solitary waves $u(t,y,z)$ is directly given through Eq. (\ref{NLSEKP}),
exploiting the soliton expression for $\eta(\tau,\upsilon,\varsigma)$  in Eq. (\ref{2kpsT}). 
Figure \ref{f3}a shows the spatio-temporal envelope intensity profile
$|u|^2$ of a (2+1)D NLSE dark X solitary wave in the $(y,t')$ plane at $z=0$, while the same profile at $z=10$ is shown Fig. \ref{f3}b.  
Here,  $t'=t-c_0z$ stands for the retarded time in the frame where the solitary wave is stationary, and we set $\epsilon_1=0.31$ and $\epsilon_2=0.17$. 

%
In particular, Fig. \ref{f3}a illustrates an exact X shape, formed at the intersection point on the origin, which is given by the sum of the dark line solitary waves. 
The solution has, asymptotically, two line dark waves for $t\ll 0$ and two for $t\gg 0$, with intensity depth $\frac{1}{2} (\epsilon_1+\epsilon_2)^2$ and angles of $\pm {\rm tan}^{-1}(\epsilon_1-\epsilon_2)$, 
measured  from the $y$ axis. 
Upon propagation, the initial X shape experiences a sort of a fission and generates a large
amplitude solitary notch at the intersection point. The observed amplification of the notch means
that the initial waveform, which is given by the sum of two dark line solitary waves, creates a large dispersive perturbation
at the intersection point, that opens up in a resonant solitary box. 
\bc{Because of this distortion, the T-type dark X solitary wave cannot be considered as a strictly invariant wave.
In order to check whether this behavior is fully reproduced in the NLSE dynamics, we report  in Fig. \ref{f3}c the outcome of the numerical integration of the NLSE at $z=10$.
By comparing Fig. \ref{f3}b and Fig. \ref{f3}c, we conclude that T-type analytical solutions provides an excellent approximation of the dynamics ruled by the hyperbolic NLSE.}

Overall, these results provide a clear evidence that  theoretical and experimental
phenomenologies of the hydrodynamic shallow water X waves dynamics can be mapped
into the realm of multidimensional spatio-temporal nonlinear optics.


Let us finally discuss the important issue of the stability of the predicted dark X solitary waves of the hyperbolic NLSE.  
Two instability factors may affect the propagation of these waves. The first one is the modulation instability (MI) of the continuous wave background.
In the case considered here ($\alpha <0$, $\beta, \gamma>0$) MI is of the conical type 
\cite{yuen80,newton,luther94}. Generally speaking, MI can be advantageous to form X waves from completely 
different initial conditions both in the absence \cite{conti03} or in the presence \cite{komi05} of the background. 
The second mechanism is related to the transverse instability of the line solitons that compose the asymptotic state of the X wave \cite{rypdal89}.
We point out that such instability is known to occur for the NLSE despite the fact that line solitons are transversally stable in the framework of 
the KP-II (unlike those of the KP-I) \cite{KP70}. However, in our simulations of the NLSE (we performed different runs for other values of the parameters), these transverse instabilities
never appears, since they are extremely long range especially for shallow solitons. \bc{In fact, we found that the primary mechanism that affects the stability of dark X solitary waves  is the MI of the background.
As a result the onset of MI causes the distortion of the solitary waves due to the amplification of spatio-temporal frequencies which are outside the spatio-temporal soliton spectrum. 
However, typically this occurs only after tens of nonlinear lengths, usually beyond the sample lengths employed in optical experiments.
Indeed, the effect of MI becomes visible only for distances longer than those shown in Figs. 1-2, i.e. for  $z>10-20$, which correspond to real-world distances $Z > 30-60 L_{nl}$.}

In summary, we have predicted the existence of optical spatiotemporal dark X solitary waves in media described by the (2+1)D hyperbolic NLSE, ruling the propagation in self-focusing and normally dispersive media. 
In particular, we have shown, analytically and numerically, families of optical dark X solitary waves of the NLSE, derived from families of  shallow water X wave solitons of the KP-II model. 
This finding opens a novel path for the excitation and control of X waves in nonlinear optics and in other areas where such NLSE apply (Bose-Einstein condensation, acoustics);
in fact, the nonlinear dark X solitary wave solutions of the NLSE are potentially observable in the regimes investigated experimentally in \cite{ditra03,faccio06,silb07} and also in \cite{eise01, baro04, baro06}. 
\bc{Future work will be devoted to investigate the possibility to excite wavepackets with similar form, starting from suitable initial conditions that differ from the strict X-shaped nonlinear modes.}

\end{document}